\documentclass[aps,orb,amssymb,showpacs,prd]{revtex4-2}
\usepackage{graphicx}
\usepackage{color}
\usepackage{dcolumn}
 \usepackage{amsmath}
\usepackage{color}
\begin{document}

\def\bc{\textbf{c}}
\def\bV{\textbf{V}}
\def\bg{\textbf{g}}
\def\bk{\textbf{k}}
\def\be{\begin{equation}}
\def\ee#1{\label{#1}\end{equation}}
\def\d{\textsf{d} }
\def\b{\textsf{b} }
\def\bu{\textbf{u}}
\def\bp{\textbf{p}}
\def\bx{\textbf{x}}
\def\bw{\textbf{w}}
\def\bv{\textbf{v}}

\newcommand{\ben}{\begin{eqnarray}}
\newcommand{\een}{\end{eqnarray}}
\def\lb{\label}
\def\no{\nonumber}

\title{Fokker-Planck equation for the Brownian motion in the post-Newtonian approximation}

\author{Gilberto M. Kremer}\email{kremer@fisica.ufpr.br}\affiliation{Universidade Federal do Paran\'a, Curitiba, Brazil}

\begin{abstract}

A mixture of light-gas particles and Brownian heavy particles is analyzed within the framework of a post-Newtonian Boltzmann equation to determine the Fokker-Planck equation for the Brownian motion. For each species, the equilibrium distribution function refers to the corresponding post-Newtonian  Maxwell-J\"uttner distribution function. The expressions for the friction viscous coefficient in the first and second post-Newtonian approximations are determined, and we show their dependence on the corresponding gravitational potentials. A linear stability analysis in the Newtonian and post-Newtonian Fokker-Planck equations for the Brownian motion is developed, where the perturbations are assumed to be  plane harmonic waves of small amplitudes. From a dispersion relation it follows that: (i) for perturbation wavelengths smaller than the Jeans wavelength two propagating modes -- corresponding to harmonic waves in opposite directions -- and one  mode that does not propagate show up; (ii) for perturbation wavelengths bigger than the Jeans wavelength the time evolution of the perturbation corresponds to a growth or a decay and the one which grows refers to the instability. 
\end{abstract}

 \maketitle

\section{Introduction}\lb{sec1}

The determination of the Fokker-Planck equation for the Brownian motion is an old topic in the literature. Among the pioneer authors, we quote Green \cite{Green} who established a generalized Fokker-Planck equation
for the process from a classical method proposed by Lord Rayleigh, Chandrasekhar \cite{Ch0} who applied stochastic analysis in the derivation of the Fokker-Planck equation, and Wang-Chang and Uhlenbeck \cite{WCU} who derived the Fokker-Planck equation from the Boltzmann equation. The applications of Brownian theory include several sciences like physics, astronomy, mathematics, chemistry, biology, economy, and some other aspects of the social and life sciences.

The Fokker-Planck equation for a single gas and for a mixture consisting  of  non-relativistic Brownian particles and a relativistic perfect gas of light particles was derived from a relativistic Boltzmann equation in the paper \cite{Guilh}. Based on the covariant Fokker-Planck equation for a single gas derived in \cite{Guilh}, a post-Newtonian version of the Fokker-Planck equation \cite{jav} was determined.

The aim of the present work is to analyze a mixture of Brownian particles and a  perfect gas of light particles within the framework of the post-Newtonian Boltzmann equation \cite{Rez,Ped,PGMK,GK}.

The starting point is the post-Newtonian Boltzmann equation for the Brownian particles, where its collision term  takes into account only the collisions between the light-gas  particles and the Brownian particles, due to the assumption that the particle number density of the Brownian particles is small with respect to that of the gas. Furthermore, the rest mass of the light-gas particles is considered much smaller than those of the Brownian particles. The equilibrium distribution functions are ruled by post-Newtonian Maxwell-J\" uttner distribution functions. By following the same strategy of Wang-Chang and Uhlenbeck \cite{WCU} the post-Newtonian Fokker-Planck equation for the Brownian motion is derived. The friction viscous coefficient obtained depends on the gravitational potentials of the post-Newtonian theory. In the non-relativistic limiting case, the Fokker-Planck equation and the friction viscous coefficient derived in \cite{WCU} are recovered. Without  dependence on the gravitational potentials,  the  relativistic expression for the friction viscous coefficient matches the one given in \cite{Guilh}. A linear stability analysis of the Newtonian and post-Newtonian Fokker-Planck equations is investigated, where the perturbations from a background state of the distribution function and gravitational potentials  are considered  as plane waves of small amplitudes. From a dispersion relation and by  introducing  the so-called Jeans wavelength,  it is shown that  perturbation wavelengths  smaller than Jeans wavelength imply a damped harmonic  wave, while  those that are greater than Jeans wavelength  the amplitude of the perturbation  grows  or decays; the one that grows refers to an instability.

The paper is structured as follows: In Section \ref{sec2} the basic equations of the post-Newtonian theory are introduced. The derivation of the post-Newtonian Fokker-Planck equation for the Brownian motion is the subject of Section \ref{sec3}. Section \ref{sec4} is dedicated to the linear stability analysis of the Newtonian and post-Newtonian  Fokker-Planck equation. Conclusions are given in the last section of the paper.

\section{Basic equations}\lb{sec2}
The Boltzmann equation is an integro-differential equation for the space-time evolution of the  one-particle distribution function in the phase space spanned by the space-time coordinates $(\bx,t)$ and  momentum of the particles $\bp$. For a relativistic gas in gravitational fields composed by $n$ species  the Boltzmann equation for the one-particle distribution function $f(\bx,\bp_a,t)\equiv f_a$ of species $a$ reads \cite{CK}
\ben\lb{brw1}
 p_a^\mu\frac{\partial f_a}{
\partial x^\mu}-\Gamma_{\mu\nu}^ip_a^\mu p_a^\nu\frac{\partial f_a}{\partial p_a^i}=
\sum_{b=1}^n\int\left(f_a'f_b'-f_bf_a\right) \,F_{ab}\,\sigma_{ab}\,d\Omega
\sqrt{- g}\frac{d^3p_b}{p_{b0}}\equiv \mathcal{Q}(f_a,f_b).
\een
The  momentum of species $a$ is $p_a^\mu$, where $p_a^\mu p_{a\mu}=m_ac^2$ and $m_a$ is the particle rest mass. $\Gamma_{\mu\nu}^i$ denote Christoffel symbols and the collision operator $ \mathcal{Q}(f_a,f_b)$ represents the binary collisions of the particle species $a$ and $b$  which before collision have momenta $(\bp_a,\bp_b)$ and after collision $(\bp'_a,\bp'_b)$. $F_{ab}=\sqrt{(p_a^\mu p_{b\mu})^2-m_a^2m_b^2c^4}$ is the invariant flux and $d\Omega=\sin\chi d\chi d\epsilon$ the element of solid angle  of the scattering process where $\chi$ represents the scattering angle and $\epsilon$ the azimuthal angle. $\sigma_{ab}$ denotes the invariant differential cross-section, $\sqrt{- g}$ the determinant of the metric tensor ${g}_{\mu\nu}$ and $f_a'\equiv f(\bx,\bp'_a,t)$, $f_b'\equiv f(\bx,\bp'_b,t)$. 

For a single gas, the collisionless Boltzmann equation in the post-Newtonian approximation  was derived in the works \cite{Rez,Ped,PGMK,GK}. For a mixture of $n$ species where the collisions of the particle species are taken into account the Boltzmann equation for species $a$ in the first post-Newtonian approximation becomes
\ben\no
&&\bigg[\frac{\partial f_a}{\partial t}+v^a_i\frac{\partial f}{\partial x^i}+\frac{\partial f_a}{\partial v_a^i}\frac{\partial U}{\partial x^i}\bigg]\bigg[1+\frac1{c^2}\left(\frac{v_a^2}2+U\right)\bigg]+\frac1{c^2}\frac{\partial f_a}{\partial v_a^i}\bigg\{v_a^2\frac{\partial U}{\partial x^i}-4v^a_iv^a_j\frac{\partial U}{\partial x^j}-4U\frac{\partial U}{\partial x^i}
\\\lb{brw2}
&&
-3v^a_i\frac{\partial U}{\partial t}+\frac{\partial\Pi_i}{\partial t}
+2\frac{\partial\Phi}{\partial x^i}+v^a_j\bigg(\frac{\partial\Pi_i}{\partial x^j}-\frac{\partial\Pi_j}{\partial x^i}\bigg)
\bigg\}=\frac1{m_a}\mathcal{Q}(f_a,f_b).
\een
As usual in the theory of mixtures the Boltzmann equation for each species has the same structure as the one for a single species, where the  mixture characteristic comes from the collision term, since it takes into account the collisions between different species. Here the left-hand side of (\ref{brw2}) has the same structure of the Boltzmann equation for a single species.
In the first post-Newtonian approximation apart from the Newtonian gravitational potential $U$, a scalar gravitational potential  $\Phi$ and a vector gravitational potential  $\Pi_i$ appear. The gravitational potentials  satisfy the   Poisson equations \cite{Ch1,Wein,GK,GK2}
\ben\lb{m2a}
\nabla^2U=-\frac{4\pi G}{c^2} \sum_{a=1}^n{\buildrel\!\!\!\! _{0} \over{T_a^{00}}},
\qquad \nabla^2\Phi=-2\pi G \sum_{a=1}^n\left({\buildrel\!\!\!\! _{2} \over{T_a^{00}}}+{\buildrel\!\!\!\! _{2} \over{T_a^{ii}}}\right),
\qquad\nabla^2\Pi^i=-\frac{16\pi G}{c}\sum_{a=1}^n{\buildrel\!\!\!\! _{1} \over{T_a^{0i}}}+\frac{\partial^2U}{\partial t\partial x^i}.
\een
Here ${\buildrel\!\!\!\! _{n} \over{T_a^{\mu\nu}}}$ denotes the  $1/c^n$-order  of the energy-momentum tensor of species $a$ and $G$ the universal gravitational constant. The definition of the energy-momentum tensor of species $a$ is given  in terms of the one-particle distribution function $f_a$ and of the particle four-velocity  $u_a^\mu$, by \cite{CK,GK}
\ben\lb{m3a}
T_a^{\mu\nu}=m_a^4c\int u_a^\mu u_a^\nu f_a\frac{\sqrt{-g}\,d^3 u_a}{u^a_0}.
\een

In the first post-Newtonian approximation, the four-velocity components $u_a^\mu$ of the species $a$  is given by  \cite{Ch1,Wein,GK}
\ben\lb{brw4a}
u_a^0=c\left[1+\frac1{c^2}\left(\frac{v_a^2}2+U\right)+\frac1{c^4}\left(\frac{3v_a^4}8+\frac{5Uv_a^2}2+\frac{U^2}2+2\Phi-\Pi_iv^a_i\right)\right],\qquad u_a^i=\frac{u_a^0v_a^i}c,
\een
while the components of the metric tensor read
\ben\lb{brw4b}
{g}_{00}=1-\frac{2U}{c^2}+\frac2{c^4}\left(U^2-2\Phi\right),\qquad {g}_{0i}=\frac{\Pi_i}{c^3},\qquad {g}_{ij}=-\left(1+\frac{2U}{c^2}\right)\delta_{ij}.
\een

From (\ref{brw4a}) and (\ref{brw4b}) one can evaluate the   determinant of the metric tensor $\sqrt{- g}$, the invariant flux  $F_{ab}$ and the element of integration  ${d^3p_b}/{p_{b0}}$ in the first post-Newtonian approximation, yielding
\ben\lb{brw5a}
&&\sqrt{-g}=1+\frac{2U}{c^2},\qquad \frac{d^3p_b}{p_{b0}}=m_b^2\left[1+\frac1{c^2}\left(2v_b^2+4U\right)\right]\frac{d^3v_b}c,
\\\no
&&F_{ab}=m_am_b\sqrt{({\rm g}_{\mu\nu}u_a^\mu u_b^\nu)^2 -c^4}=c\,m_am_b\,\sqrt{v_a^2+v_b^2-2\bv_a\cdot\bv_b}
\\\lb{brw5b}
&&\qquad\times\left\{1+\frac1{c^2}\left[2U+\frac{v_a^4+v_b^4+v_a^2v_b^2-2(v_a^2+v_b^2)\bv_a\cdot\bv_b+(\bv_a\cdot\bv_b)^2}{2(v_a^2+v_b^2-2\bv_a\cdot\bv_b)}\right]\right\}.
\een

\section{Post-Newtonian equation for the Brownian particle}\lb{sec3}

Let us consider a mixture of two species where one of them is characterized by a perfect gas of light-rest mass particles $m_G$, while the other refers to Brownian particles with heavy-rest mass $m_B\gg m_G$, and that the particle number density of the Brownian particles is much smaller than the one of the perfect gas, i.e. $n_B/n_G\ll1$. These premises about Brownian particles  suggest that 
the collision term between the Brownian particles can be neglected and the Boltzmann equation (\ref{brw2}) for the one-particle distribution function of the Brownian particles reduces to 
\ben\no
&&\bigg[\frac{\partial f_B}{\partial t}+v^B_i\frac{\partial f}{\partial x^i}+\frac{\partial f_B}{\partial v_B^i}\frac{\partial U}{\partial x^i}\bigg]\bigg[1+\frac1{c^2}\left(\frac{v_B^2}2+U\right)\bigg]+\frac1{c^2}\frac{\partial f_B}{\partial v_B^i}\bigg\{v_B^2\frac{\partial U}{\partial x^i}-4v^B_iv^B_j\frac{\partial U}{\partial x^j}-4U\frac{\partial U}{\partial x^i}-3v^B_i\frac{\partial U}{\partial t}+\frac{\partial\Pi_i}{\partial t}
\\\lb{brw7}
&&
+2\frac{\partial\Phi}{\partial x^i}+v^B_j\bigg(\frac{\partial\Pi_i}{\partial x^j}-\frac{\partial\Pi_j}{\partial x^i}\bigg)
\bigg\}=\frac1{m_B}\int\left(f_B'f_G'-f_Bf_G\right) \,F\,\sigma\,d\Omega
\sqrt{-g}\frac{d^3p_G}{p_{G0}}=\frac1{m_B}\mathcal{Q}(f_B,f_G).
\een
For simplicity we have written $F\equiv F_{BG}$ and $\sigma\equiv \sigma_{BG}$.

 For a gas described by the Boltzmann equation, the equilibrium state is  characterized by a vanishing collision term, where the number of particles that enter the phase space volume is equal to those that leaving it. For mixture of relativistic gases the one-particle distribution function for  species $a$ is given by the Maxwell-J\"uttner distribution function
 \ben\lb{sd}
f_a(\bx,\bp_a,t)=\frac{n_a}{4\pi m_a^2ckTK_2(m_ac^2/kT)}\exp\left(-\frac{p^\mu_aU_{a\mu}}{kT}\right).
 \een
 Here $k$ denotes the Boltzmann constant, $T$ the absolute temperature of the gas species and $\zeta_a=m_ac^2/kT$ is a relativistic parameter representing the ratio of the rest energy of the gas particles $m_ac^2$ and the thermal energy of the gas species $kT$. In the non-relativistic regime $m_ac^2/kT\gg1$ and in the ultra-relativistic regime  $m_ac^2/kT\ll1$. Furthermore, $K_2(m_ac^2/kT)$ represents a modified Bessel function of second kind.

We assume that the gas is at equilibrium where its one-particle distribution function is characterized by the post-Newtonian Maxwell-J\"uttner distribution function \cite{PGMK,GK} which follows from (\ref{sd}), namely
\ben\lb{brw8a}
 f_G\vert_E=f^G_{MJ}=f_G^{(0)}\left\{1-\frac1{\zeta_G}\left[\frac{15}8+2\frac{m_G^2 v_G^2}{k^2T^2}\left(U+\frac{3v_G^2}{16}\right)\right]\right\},\qquad
f_G^{(0)}=\frac{n_G}{(2\pi m_GkT)^\frac32}e^{-\frac{m_Gv_G^2}{2kT}}.
\een

The collisions of  Brownian particles with gas particles affect only a small deviation from equilibrium of the former so that we can represent the Brownian one-particle distribution function as $f_B=f_B\vert_Eh(\bv_B)$ where
\ben\lb{brw8b}
  f_B\vert_E=f^B_{MJ}=f_B^{(0)}\left\{1-\frac1{\zeta_B}\left[\frac{15}8+2\frac{m_B^2 v_B^2}{k^2T^2}\left(U+\frac{3v_B^2}{16}\right)\right]\right\},\qquad
f_B^{(0)}=\frac{n_B}{(2\pi m_BkT)^\frac32}e^{-\frac{m_Bv_B^2}{2kT}}.
\een
In the above equation, $h(\bv_B)$ is the equilibrium deviation of the one-particle distribution function for the Brownian particle  when space homogeneity is assumed and  $\zeta_B=m_Bc^2/kT$. We can expand the equilibrium deviation for  the post-collision velocity $h(\bv_B^{\prime})$ in the following Taylor series 
\ben\lb{brw9}
h(\bv_B^{\prime})\approx h(\bv_{B})+\Delta v_i^B\frac{\partial h}{\partial v_B^i}+\frac12\Delta v_i^B\Delta v_j^B\frac{\partial^2 h}{\partial v_B^i\partial v_B^j},
\een
by retaining up to second terms in $\Delta v_i^B=v_i^{B\prime}-v_i^{B}$. 

The collision term of the Brownian Boltzmann equation (\ref{brw7}) together with (\ref{brw8a}) -- (\ref{brw9}) can be written as
\ben \lb{brw10a}
\frac1{m_B}\mathcal{Q}(f_B,f_G)= \frac{f_{MJ}^B}{m_B}\mathcal{I}(h),
\een
where $\mathcal{I}(h)$ is the following integral 
\ben\lb{brw10b}
\mathcal{I}(h)=\int\left[\Delta v_i^B\frac{\partial h}{\partial v_B^i}+\frac12\Delta v_i^B\Delta v_j^B\frac{\partial^2 h}{\partial v_B^i\partial v_B^j}\right]f^G_{MJ}\,F\,\sigma\,d\Omega
\sqrt{- g}\frac{d^3p_G}{p_{G0}}.
\een
Note that  $f_{MJ}^{B'}f_{MJ}^{G'}=f_{MJ}^{B}f_{MJ}^{G}$ thanks to  energy-momentum conservation laws. 

For the evaluation of the integral $\mathcal I$ we introduce the relative velocities of the species before and after collision, namely
\ben\lb{brw11a}
\Upsilon_i=\frac{c u_G^{ i}}{u_G^{ 0}}-\frac{c u_B^{ i}}{u_B^{ 0}}=v_i^{G}-v_i^{B},\qquad
\Upsilon^\prime_i=\frac{c u_G^{\prime i}}{u_G^{\prime 0}}-\frac{c u_B^{\prime i}}{u_B^{\prime 0}}=v_i^{G\prime}-v_i^{B\prime}.
\een

If we take into account the energy and momentum conservation laws and  the approximation $m_G/m_B\ll1$ we obtain from (\ref{brw11a}) that $\vert{\bf\Upsilon}^\prime\vert\approx\vert{\bf \Upsilon}\vert$. Furthermore, we have 
\ben\lb{brw11b}
\Delta \Upsilon_i=\Upsilon_i^\prime-\Upsilon_i=-\left\{1+\frac{m_B}{m_G}\left[1-\frac1{c^2}\left(\frac{v_G^2}{2}+U\right)\right]\right\}\left(v_i^{B\prime}-v_i^B\right)\approx-\frac{m_B}{m_G}\left[1-\frac1{c^2}\left(\frac{v_G^2}{2}+U\right)\right]\Delta v_i^B.
\een

 From the relationship $\vert{\bf\Upsilon}^\prime\vert\approx\vert{\bf \Upsilon}\vert$ between the moduli of the relative velocities before and after collision  and considering  the scattering angle $\chi$ and the azimuthal angle $\epsilon$ as the spherical angles of $\Upsilon_i'$ with respect to $\Upsilon_i$ one can perform the integration in the azimuthal angle of the relative velocities  difference  $\Delta \Upsilon_i$, yielding
 \ben\lb{brw12a}
&&\int_0^{2\pi}d\epsilon\,\Delta \Upsilon_i=2\pi(\cos\chi-1)\left(v_i^G-v_i^B\right),
\\\lb{brw12b}
&&\int_0^{2\pi}d\epsilon\,\Delta \Upsilon_i\Delta \Upsilon_j=2\pi\bigg\{\left[(1-\cos\chi)^2-\frac12\sin^2\chi\right]\left(v_i^G-v_i^B\right)\left(v_j^G-v_j^B\right)+\frac{(v_G^2+v_B^2-2v_k^Gv_k^B)}2\sin^2\chi\delta_{ij}\bigg\}.\quad
 \een
Furthermore, let  $\Theta$ and $\Phi$ be the  polar angles of the vector $v_i^G$  with with respect to the vector $v_i^B$, then the integration in the angle $\Phi$ of $\Delta v_i^B$ and  $\Delta v_i^B\Delta v_j^B$ can be carry out, resulting
\ben\lb{brw13a}
&&\int_0^{2\pi}\int_0^{2\pi}d\epsilon\,d\Phi\,\Delta v_i^B=(2\pi)^2\frac{m_G}{m_B}\left[1+\frac1{c^2}\left(\frac{v_G^2}{2}+U\right)\right](1-\cos\chi)\left[\frac{v_G}{v_B}\cos\Theta-1\right]v_i^B,
\\\no
&&\int_0^{2\pi}\int_0^{2\pi}d\epsilon\,d\Phi\,\Delta v_i^B\Delta v_j^B=(2\pi)^2\frac{m_G^2}{m_B^2}\left[1+\frac1{c^2}\left(v_G^2+2U\right)\right]\bigg\{\bigg((1-\cos\chi)^2-\frac12\sin^2\chi\bigg)\bigg[\frac{v_G^2}{2}\sin^2\Theta\delta_{ij}
\\\lb{brw13b}
&&\qquad+v_i^Bv_j^B\bigg(\bigg(\frac{v_G}{v_B}\cos\Theta-1\bigg)^2-\frac{v_G^2}{2v_B^2}\sin^2\Theta\bigg)\bigg]+\pi(v_G^2+v_B^2-2v_Gv_B\cos\Theta)\sin^2\chi\delta_{ij}\bigg\}.
\een

The gas integration element follows from (\ref{brw5a}) and can be written as
\ben\lb{brw14a}
\sqrt{-g}\frac{d^3p_G}{p_{G0}}=m_G^2\left[1+\frac1{c^2}\left(2v_G^2+6U\right)\right]\frac{v_G^2\sin\Theta d\Theta d\Phi dv_G}c,
\een
once the spherical coordinates of the gas velocity $(v_G,\Theta,\Phi)$ are introduced.

Now we introduce dimensionless particle velocities $\mathcal{V}_i^B, \mathcal{V}_i^G$ and the dimensionless Newtonian potential $\mathcal{U}$:
\ben\lb{brw15}
\mathcal{V}_i^B=v_i^B\sqrt{\frac{m_B}{2kT}},\qquad \mathcal{V}_i^G=v_i^G\sqrt{\frac{m_G}{2kT}},\qquad \mathcal{U}=\frac{m_GU}{kT}.
\een

By retaining terms up to the order $\mathcal{O}(\sqrt{m_G/m_B})$, the invariant flux (\ref{brw5b})  
  becomes
\ben\lb{brw14b}
F=c \sqrt{\frac{2kT}{m_G}}m_Bm_G \mathcal{V}_G\sqrt{1-2\sqrt{\frac{m_G}{m_B}}\frac{\mathcal{V}_B}{\mathcal{V}_G} \cos\Theta}\left\{1+\frac1{\zeta_G}\left[2\mathcal{U}+{\mathcal{V}_G^2}\right]\right\}.
\een
Note  that in the above expression it was also considered that   $v_B /v_G \simeq\sqrt{m_G/m_B}$, which 
 is required
for a linear dependence of the viscosity coefficient on the velocity (see \cite{WCU}).

The product of the gas integration element (\ref{brw14a}) with the invariant flux (\ref{brw14b}), yields
\ben\no
&&F\sqrt{- g}\frac{d^3p_G}{p_{G0}}=m_Bm_G^3\mathcal{V}_G^3\left(\frac{2kT}{m_G}\right)^2\sqrt{1-2\sqrt{\frac{m_G}{m_B}}\frac{\mathcal{V}_B}{\mathcal{V}_G}\cos\Theta}\left\{1+\frac1{\zeta_G}\left[8\mathcal{U}+5\mathcal{V}_G^2\right]\right\}\sin\Theta d\Theta d\Phi d\mathcal{V}_G
\\\lb{brw16a}
&&\qquad\approx m_Bm_G^3\mathcal{V}_G^3\left(\frac{2kT}{m_G}\right)^2\left(1-\sqrt{\frac{m_G}{m_B}}\frac{\mathcal{V}_B}{\mathcal{V}_G}\cos\Theta\right)\left\{1+\frac1{\zeta_G}\left[8\mathcal{U}+5\mathcal{V}_G^2\right]\right\}\sin\Theta d\Theta d\Phi d\mathcal{V}_G.
\een

The differential cross section is a function of the scattering angle $\chi$ and the invariant flux $F$. Hence,  up to the order  $\mathcal{O}(\sqrt{m_G/m_B})$  the following approximation holds
\ben\lb{brw16b}
\sigma(\chi,F)\approx \sigma\left(\chi,\sqrt{\frac{2kT}{m_G}}\mathcal{V}_G\right)\left\{1+\frac{\partial\ln\sigma}{\partial \ln\mathcal{V}_G}\left[\frac1{\zeta_G}\left(2\mathcal{U}+\mathcal{V}_G^2\right)-\sqrt{\frac{m_G}{m_B}}\frac{\mathcal{V}_B}{\mathcal{V}_G}\cos\Theta
\right]\right\}.
\een

The integrals (\ref{brw13a}) and (\ref{brw13b}) by considering the terms up to the order  $\mathcal{O}(\sqrt{m_G/m_B})$ read
\ben\no
&&\int_0^{2\pi}\int_0^{2\pi}d\epsilon\,d\Phi\left\{\Delta v_i^B\frac{\partial h}{\partial v_B^i}+\frac12\Delta v_i^B\Delta v_j^B\frac{\partial^2 h}{\partial v_B^i\partial v_B^j}\right\}=(2\pi)^2\frac{m_G}{m_B}\left[1+\frac1{\zeta_G}\left(\mathcal{V}_G^2+U)\right)\right](1-\cos\chi)
\\\no
&&\qquad\times\bigg\{\mathcal{V}_i^B\frac{\partial h}{\partial \mathcal{V}^i_B}\left(\sqrt{\frac{m_B}{m_G}}\frac{\mathcal{V}_G}{\mathcal{V}_B}\cos\Theta-1\right)-\left[1+\frac1{\zeta_G}\left(\mathcal{V}_G^2+U)\right)\right]\bigg[\frac18\mathcal{V}_i^B\mathcal{V}_j^B\frac{\partial^2 h}{\partial \mathcal{V}^i_B\partial \mathcal{V}^j_B}(3\cos^2\Theta-1)(3\cos\chi-1)
\\\lb{brw16c}
&&\qquad-\frac18\mathcal{V}_G^2\frac{\partial^2 h}{\partial \mathcal{V}^i_B\partial \mathcal{V}^i_B}\left[(3\cos^2\Theta-1)\cos\chi+3-\cos^2\Theta\right]\bigg]\bigg\}.
\een

We collect the partial results (\ref{brw16a}) -- (\ref{brw16c}) and together with the expression for the Maxwell-J\"uttner distribution function for the gas species (\ref{brw8a}) we insert into (\ref{brw10b})  and get
\ben\no
&&\mathcal{I}(h)=(2\pi)^2m_G n_G\sqrt{\frac{2kT}{\pi^3m_G}}\int_0^\pi d\chi\sin\chi(1-\cos\chi)\int_0^\infty d\mathcal{V}_G \mathcal{V}_G^3e^{-\mathcal{V}_G^2}\bigg\{1-\frac1{\zeta_G}\bigg[\frac{15}8+4\mathcal{V}_G^2\bigg(\mathcal{U}+\frac{3\mathcal{V}_G^2}{8}\bigg)-8\mathcal{U}-5\mathcal{V}_G^2\bigg]\bigg\}
\\\no
&&\qquad\times\int_0^\pi d\Theta\sin\Theta \,\sigma\left(\chi,\sqrt{\frac{2kT}{m_G}}\mathcal{V}_G\right)\left[1+\frac1{\zeta_G}\left(2\mathcal{U}+\mathcal{V}_G^2\right)\frac{\partial\ln\sigma}{\partial \ln\mathcal{V}_G}-\sqrt{\frac{m_G}{m_B}}\frac{\mathcal{V}_B}{\mathcal{V}_G}\cos\Theta\left(1+\frac{\partial\ln\sigma}{\partial \ln\mathcal{V}_G}\right)\right]
\\\no
&&\qquad
\times\left[1+\frac1{\zeta_G}\left(\mathcal{U}+\mathcal{V}_G^2\right)\right]\bigg\{\mathcal{V}_i^B\frac{\partial h}{\partial \mathcal{V}^i_B}\left(\sqrt{\frac{m_B}{m_G}}\frac{\mathcal{V}_G}{\mathcal{V}_B}\cos\Theta-1\right)-\left[1+\frac1{\zeta_G}\left(\mathcal{V}_G^2+U)\right)\right]\bigg[\frac18\mathcal{V}_i^B\mathcal{V}_j^B\frac{\partial^2 h}{\partial \mathcal{V}^i_B\partial \mathcal{V}^j_B}
\\\lb{brw17a}
&&\qquad\times(3\cos^2\Theta-1)(3\cos\chi-1)-\frac18\mathcal{V}_G^2\frac{\partial^2 h}{\partial \mathcal{V}^i_B\partial \mathcal{V}^i_B}\left[(3\cos^2\Theta-1)\cos\chi+3-\cos^2\Theta\right]\bigg]\bigg\}.
\een
 Note that for a non-relativistic gas species $\zeta_G\gg1$ and (\ref{brw17a}) reduce  to the equation for $I(h)$ given on page 91 of the work by Wang-Chang and Uhlenbeck \cite{WCU}.

The $\Theta$--integral in (\ref{brw17a}) can be performed, yielding
\ben\no
&&\mathcal{I}(h)=\frac83\pi^2m_G n_G\sqrt{\frac{2kT}{\pi^3m_G}}\int_0^\pi d\chi\sin\chi(1-\cos\chi)\int_0^\infty d\mathcal{V}_G \mathcal{V}_G^3e^{-\mathcal{V}_G^2}\sigma\left(\chi,\sqrt{\frac{2kT}{m_G}}\mathcal{V}_G\right)
\\\no
&&\qquad\times\bigg\{1-\frac1{\zeta_G}\bigg[\frac{15}8+4\mathcal{V}_G^2\bigg(\mathcal{U}+\frac{3\mathcal{V}_G^2}{8}\bigg)-10\mathcal{U}-7\mathcal{V}_G^2\bigg]\bigg\}\bigg\{\mathcal{V}_G^2\frac{\partial^2 h}{\partial \mathcal{V}^i_B\partial \mathcal{V}^i_B}\left[1+\frac1{\zeta_G}\left(2\mathcal{U}+\mathcal{V}_G^2\right)\frac{\partial\ln\sigma}{\partial \ln\mathcal{V}_G}\right]
\\\lb{brw17b}
&&\qquad-\mathcal{V}_i^B\frac{\partial h}{\partial \mathcal{V}^i_B}\left[4+\frac{\partial\ln\sigma}{\partial \ln\mathcal{V}_G}\right]\bigg\}.
\een

Now considering the partial integration
\ben\lb{brw17c}
&&\int_0^\infty d\mathcal{V}_G \mathcal{V}_G^4e^{-\mathcal{V}_G^2}\frac{\partial\sigma}{\partial\mathcal{V}_G}=\int_0^\infty d\mathcal{V}_G \,\sigma\,\mathcal{V}_G^3e^{-\mathcal{V}_G^2}\left[2\mathcal{V}_G^2
-4\right],
\\
&&\int_0^\infty d\mathcal{V}_G \mathcal{V}_G^6e^{-\mathcal{V}_G^2}\frac1{\zeta_G}\left(2\mathcal{U}+\mathcal{V}_G^2\right)\frac{\partial\sigma}{\partial \mathcal{V}_G}=\int_0^\infty d\mathcal{V}_G \,\sigma\,\mathcal{V}_G^5e^{-\mathcal{V}_G^2}\frac1{\zeta_G}\left(4\mathcal{U}\mathcal{V}_G^2+2\mathcal{V}_G^4-12\mathcal{U}-8\mathcal{V}_G^2\right),
\een
the expression (\ref{brw17b}) reduces to 
\ben\no
&&\mathcal{I}(h)=\frac83\pi^2m_G n_G\sqrt{\frac{2kT}{\pi^3m_G}}\int_0^\pi d\chi\sin\chi(1-\cos\chi)\int_0^\infty d\mathcal{V}_G \mathcal{V}_G^5e^{-\mathcal{V}_G^2}\sigma\left(\chi,\sqrt{\frac{2kT}{m_G}}\mathcal{V}_G\right)\bigg\{\frac{\partial^2 h}{\partial \mathcal{V}^i_B\partial \mathcal{V}^i_B}
\\\lb{brw17d}
&&\qquad-2\mathcal{V}_i^B\frac{\partial h}{\partial \mathcal{V}^i_B}\bigg\}\bigg\{1-\frac1{\zeta_G}\bigg[\frac{15}8+20\mathcal{U}\mathcal{V}_G^2+\frac{23}2\mathcal{V}_G^4-4\mathcal{U}\mathcal{V}_G^4-2\mathcal{V}_G^6-7\mathcal{V}_G^2-10\mathcal{U}\bigg]\bigg\}.
\een

We can return to the dimensional variables $(\bv_G, \bv_B, U)$ and obtain from (\ref{brw7}), (\ref{brw10a}) and (\ref{brw17d}) the Fokker-Planck equation for the Brownian motion  in the post-Newtonian approximation 
\ben\no
&&\bigg[\frac{\partial f_B}{\partial t}+v^B_i\frac{\partial f}{\partial x^i}+\frac{\partial f_B}{\partial v_B^i}\frac{\partial U}{\partial x^i}\bigg]\bigg\{1+\frac1{c^2}\left[\frac{15kT}{8m_B}+\frac{v_B^2}2+U+\frac{2m_Bv_B^2}{kT}\left(U+\frac{3v_B^2}{16}\right)\right]\bigg\}+\frac1{c^2}\frac{\partial f_B}{\partial v_B^i}\bigg\{v_B^2\frac{\partial U}{\partial x^i}
\\\no
&&
\qquad-4v^B_iv^B_j\frac{\partial U}{\partial x^j}-4U\frac{\partial U}{\partial x^i}-3v^B_i\frac{\partial U}{\partial t}+\frac{\partial\Pi_i}{\partial t}+2\frac{\partial\Phi}{\partial x^i}+v^B_j\bigg(\frac{\partial\Pi_i}{\partial x^j}-\frac{\partial\Pi_j}{\partial x^i}\bigg)
\bigg\}=\eta\frac{\partial}{\partial v_B^i}\left[v_i^Bf^{(0)}_Bh+\frac{kT}{m_B}\frac{\partial f^{(0)}_Bh}{\partial v_B^i}\right]
\\\lb{brw18a}
&&\qquad=\eta\frac{\partial}{\partial v_B^i}\left\{v_i^Bf_B^*+\frac{kT}{m_B}\frac{\partial f_B^*}{\partial v_B^i}\right\},
\een
where $f_B^*$ denotes the following expression
\ben
f_B^*=f_B\left\{1+\frac1{\zeta_B}\left[\frac{15}8+2\frac{m_B^2 v_B^2}{k^2T^2}\left(U+\frac{3v_B^2}{16}\right)\right]\right\}.
\een
Note that on the right-hand side of (\ref{brw18a}) we have taken only the Maxwellian distribution function $f_B^{(0)}$ while the relativistic term of the Maxwell-J\"uttner distribution function was introduced in its left-hand side. The coefficient $\eta$ is the so-called  friction viscous coefficient, which in the post-Newtonian approximation is given by
\ben\no
&&\eta=\frac{16\sqrt\pi}3\frac{n_Gm_G}{m_B}\left(\frac{m}{2kT}\right)^\frac52\int_0^\pi d\chi\sin\chi(1-\cos\chi)\int_0^\infty dv_Gv_G^5 \exp\left(-\frac{m_Gv_G^2}{2kT}\right)\sigma(\chi,v_G)
\\\lb{brw18b}
&&\qquad\times\bigg\{1-\frac1{\zeta_G}\bigg[\frac{15}8+\frac{m_G^2v_G^2}{k^2T^2}\left(10U+\frac{23}8v_G^2\right)-\frac{m_G^3v_G^4}{k^3T^3}\left(U+\frac{v_G^2}4\right)-\frac{m_G}{kT}\left(\frac72v_G^2+10U\right)\bigg]\bigg\}.
\een

By taking the non-relativistic limits  $\zeta_G\gg1$ and $\zeta_B\gg1$ the Fokker-Planck equation for the Brownian motion   (\ref{brw18a})  and the friction viscous coefficient (\ref{brw18b}) reduces to the corresponding terms of the work by Wang-Chang and Uhlenbeck \cite{WCU}, namely
\ben\lb{brw18c}
&&\frac{\partial f_B}{\partial t}+v^B_i\frac{\partial f_B}{\partial x^i}+\frac{\partial U}{\partial x^i}\frac{\partial f_B}{\partial v_B^i}=\eta\frac{\partial}{\partial v_B^i}\left[v_i^Bf_B+\frac{kT}{m_B}\frac{\partial f_B}{\partial v_B^i}\right],
\\\lb{brw18d}
&&\eta=\frac{16\sqrt\pi}3\frac{n_Gm_G}{m_B}\left(\frac{m}{2kT}\right)^\frac52\int_0^\pi d\chi\sin\chi(1-\cos\chi)\int_0^\infty dv_Gv_G^5 \exp\left(-\frac{m_Gv_G^2}{2kT}\right)\sigma(\chi,v_G).
\een

Let us analyze the friction viscous coefficient for the differential cross section corresponding to the potentials of the hard sphere particles, which  is characterized by a constant differential cross section  
$\sigma=\d_G^2/4$ where $\d_G$ is the diameter of a hard sphere gas particle. In this case, the integrals in (\ref{brw18b}) can be performed, yielding
\ben\lb{brw19a}
\eta=\frac{8}3\frac{n_Gm_G\,\d_G^2}{m_B}\sqrt{\frac{2\pi kT}{m_G}}\left\{1+\frac1{\zeta_G}\left[\frac{9}8-\frac{2m_GU}{kT}\right]\right\}.
\een
Here is important to note that in the first post-Newtonian approximation the friction viscous coefficient depends on the Newtonian gravitational potential $U$. This result corroborated  those of the works \cite{Andre1,Andre2,Gil2,Gil3} where it was shown that the transport coefficients of a  rarefied gas in the presence of a gravitational field  depend on the Newtonian gravitational potential.

One can ask about the expansion of the results given here to the second post-Newtonian approximation. The second post-Newtonian theory was developed by Chandrasekhar and Nutku \cite{Ch2} and the corresponding Boltzmann equation can be found in the works \cite{GK,PGMK}. The equations for the Boltzmann equation and Maxwell-J\"uttner distribution function are very large, so  we give here only the result corresponding  to the friction viscous coefficient
\ben\lb{brw19b}
\eta=\frac{8}3\frac{n_Gm_G\,\d_G^2}{m_B}\sqrt{\frac{2\pi kT}{m_G}}\left\{1+\frac1{\zeta_G}\left[\frac{9}8-\frac{2m_GU}{kT}\right]+\frac1{\zeta_G^2}\left[\frac{9}{128}-\frac{9m_GU}{4kT}-\frac{86m_G^2U^2}{k^2T^2}+\frac{6m_G^2\Phi}{k^2T^2}+\frac{m_G^2\Psi_{kk}}{3k^2T^2}\right]\right\}.
\een
In the above equation $\Phi, \Psi_{kk}$ are two new gravitational potentials which obey Poisson-type equations. If we neglect in (\ref{brw19b}) all gravitational potentials, we get the expression for this coefficient  given in \cite{Guilh}.

\section{Linear stability analysis of the Fokker-Planck equation}\lb{sec4}
\subsection{Newtonian Fokker-Planck equation}

We shall analyze the  linear stability  of the Newtonian Fokker-Planck equation (\ref{brw18c}), which  is coupled with the Poisson equation for the Newtonian gravitational potential
\ben\lb{brw20b}
&&\nabla^2 U=-4\pi Gm_B^4\left[\int f_B d^3v_B+\frac{m_G^4}{m_B^4}\int f_Gd^3v_G\right].
\een

The Newtonian Fokker-Planck equation (\ref{brw18c}) is identically satisfied at equilibrium, provided the gradient of the Newtonian gravitational potential at equilibrium also vanishes, i.e. $\nabla U_0=0$, which is a consequence of symmetry considerations for a homogeneous system. Condition $\nabla U_0=0$ does not satisfy the Poisson equation (\ref{brw20b}), since its right-hand side refers to the mass densities of the species. This inconsistency is removed by invoking "Jeans swindle" \cite{Jeans,Coles,BT1} where the Poisson equation is valid only for  perturbed distribution functions and the gravitational potential. 

The equilibrium state is subjected to small perturbations $h_B(\bx,\bv_B,t)$, $h_G(\bx,\bv_B,t)$ and $U_1(\bx,t)$, namely
\ben\lb{brw21a}
f_B=f_B^{(0)}\left[1+h_B(\bx,\bv_B,t)\right],\qquad f_G=f_G^{(0)}\left[1+h_G(\bx,\bv_G,t)\right],\qquad U(\bx,t)=U_0+U_1(\bx,t).
\een
In this case the Poisson equation (\ref{brw20b}) for the perturbations read
\ben\lb{brw22a}
\nabla^2 U_1=-4\pi m_B^4\left[\int f_B^{(0)}h_Bd^3v_B+\frac{m_G^4}{m_B^4}\int f_G^{(0)}h_Gd^3v_G\right]\approx -4\pi m_B^4\int f_B^{(0)} h_Bd^3v_B,
\een
where we have considered the approximation $m_G/m_B\ll1$.

The insertion of the representations (\ref{brw21a}) into the Fokker-Planck equation (\ref{brw18c}) leads to
\ben\lb{brw22b}
\frac{\partial h_B}{\partial t}+v^B_i\frac{\partial h_B}{\partial x^i}-\frac{m_B}{kT}v_i^B\frac{\partial U_1}{\partial x^i}=\eta\left[\frac{kT}{m_B}\frac{\partial^2 h_B}{\partial v_B^i\partial v_B^i}-v_i^B\frac{\partial h_B}{\partial v_B^i}\right],
\een
by neglecting products of all non-linear terms in the small perturbations.

The perturbations $h_B(\bx,\bv_B,t)$ and $U_1(\bx,t)$ are represented by plane waves of wavenumber vector $\bk$ and frequency $\omega$, namely
\ben\lb{brw23a}
h_B(\bx,\bv_B,t)=\overline h(\bv)\exp\left[i\left(\bk\cdot\bx-\omega t\right)\right],\qquad U_1(\bx,t)=\overline U \exp\left[i\left(\bk\cdot\bx-\omega t\right)\right],
\een
where  $\overline h(\bv)$ and $\overline U$ are small amplitudes.

We  follow \cite{Gil4,GK} and assume that the amplitude $\overline h(\bv)$ is a linear combination of the summational invariants of the Boltzmann equation, $1, v_i^B, v_B^2$ and write
\ben\lb{brw23b}
\overline h(\bv)=A+{\bf B}\cdot \bv_B+Dv_B^2,
\een
where $A, {\bf B}, D$ are unknown constants which are  assumed to be real.

Insertion of perturbations (\ref{brw23a}) and (\ref{brw23b}) into the Poisson  (\ref{brw22a}) and  Fokker-Planck (\ref{brw18c}) equations leads to the following coupled system of equations 
\ben\lb{brw24a}
&&\kappa^2\overline U=4\pi G\left(A+3\sigma_B^2 D\right)\rho_B^0,
\\\lb{brw24b}
&&\left[A+{\bf B}\cdot \bv_B+Dv_B^2\right]\left[\omega-\bk\cdot\bv_B\right]+\bk\cdot\bv_B\frac{\overline U}{\sigma_B^2}=i\eta\left[\left(6\sigma_B^2-2v_B^2\right)D-{\bf B}\cdot\bv_B\right].
\een
In the above equations $\sigma_B$ is the dispersion (thermal) velocity  of the Brownian particles, $\rho^0_B$ its background mass density and $\kappa$ the modulus of the wavenumber vector. They are given by
\ben
\sigma_B=\sqrt{\frac{kT}{m_B}},\qquad\rho_B^0=m_B\int f_B^{(0)}d^3v_B,\qquad \kappa=\sqrt{\bk\cdot\bk}.
\een

The multiplication of  (\ref{brw24b}) by the summational invariants  $1, v_i^B, v_B^2$ and integration of the resulting equations with respect to $f_B^{(0)}d^3v_B$ we get, respectively  
\ben\lb{brw25a}
&&\omega\left(A+3\sigma_B^2D\right)-\sigma_B^2{\bf B}\cdot\bk=0,
\\\lb{brw25b}
&&\left(\omega+i\eta\right) {\bf B}\cdot\bk-\kappa^2\left(A+5\sigma^2_BD-\frac{\overline U}{\sigma_B^2}\right)=0,
\\\lb{brw25c}
&&\omega\left(3A+15\sigma_B^2D\right)-5\sigma_B^2{\bf B}\cdot\bk+12i\eta D\sigma_B^2=0.
\een

\begin{figure}[ht]
\includegraphics[width=12cm]{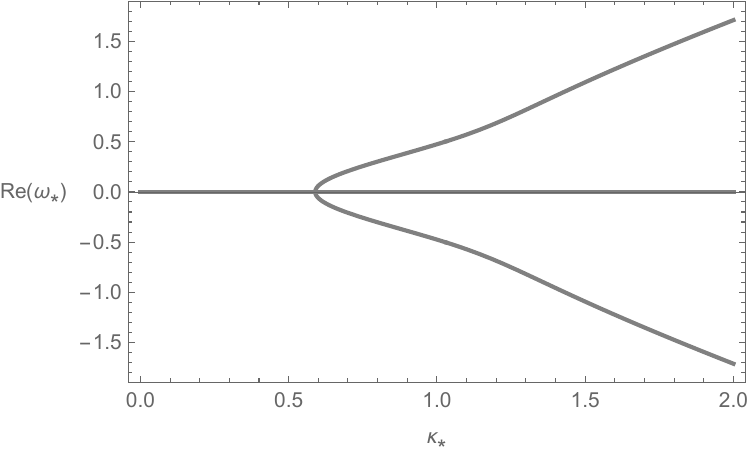}
\caption{Real part of the dimensionless frequency $\omega_*$ as function of the dimensionless wavenumber $\kappa_*$ for a dimensionless friction viscous coefficient $\eta_*=0.25$.} \label{fig1}
\end{figure}   
\unskip

Equations (\ref{brw24a}), (\ref{brw25a}), (\ref{brw25b}) and (\ref{brw25c}) become  a coupled system of algebraic equations  for the amplitudes $A, {\bf B}\cdot\bk, D, \overline U$ and a non-trivial solution of this coupled system can be found if the determinant of their  coefficients vanishes, so that it follows the dispersion relation
\ben\lb{brw26a}
\omega_*\left(\omega_*^2-\kappa_*^2+1-2\eta_*^2\right)+i\eta_*\left(2+3\omega_*^2-\frac65\kappa_*^2\right)=0.
\een
Here   the following dimensionless quantities were introduced
\ben\lb{brw26b}
\omega_*=\frac{\omega}{\sqrt{4\pi G\rho_B^0}},\qquad \eta_*=\frac{\eta}{\sqrt{4 \pi G\rho_B^0}},\qquad\kappa_*=\frac{V_s\kappa}{\sqrt{4 \pi G\rho_B^0}}=\frac{\kappa}{\kappa_J},
\een
where $\kappa_J={\sqrt{4 \pi G\rho_B^0}}/V_s$ is the Jeans wavenumber and $V_s=\sqrt{5/3}\sigma_B$ the adiabatic sound speed.

The solution of dispersion relation (\ref{brw26a}) for the dimensionless frequency $\omega_*$ has three roots given by 
\ben\lb{brw27a}
&&\omega_*=-i\eta_*-\frac{2^\frac13(1-\kappa_*^2+\eta_*^2)}{\alpha_*}+\frac{\alpha_*}{3\times2^\frac13},
\\\lb{brw27b}
&&\omega_*=-i\eta_*+\frac{(1\pm i\sqrt3)(1-\kappa_*^2+\eta_*^2)}{2^\frac23\alpha_*}-\frac{(1\mp i\sqrt3)\alpha_*}{6\times2^\frac13},
\een
where $\alpha_*$ is the following function of the dimensionless wavenumber $\kappa_*$ and friction viscous coefficient $\eta_*$
\ben\lb{brw27c}
\alpha_*=\left[-27i\eta_*+\frac{27}5i\kappa_*^2\eta_*+\sqrt{\left(-27i\eta_*+\frac{27}5i\kappa_*^2\eta_*\right)^2+4(3-3\kappa_*^2+3\eta_*^2)^3}\,\right]^\frac13.
\een

For a vanishing friction viscous coefficient, the Fokker-Planck equation (\ref{brw18c}) reduces to a collisionless Boltzmann equation for Brownian particles, and the three roots become
\ben
\omega_*=0,\qquad \omega_*=\pm\sqrt{\kappa_*^2-1}
\een
which is the Jeans solution \cite{Jeans}. Since $\kappa_*=\kappa/\kappa_J=\lambda_J/\lambda$ where $\lambda=2\pi/\kappa$ is the wavelength of the perturbation, one may conclude that perturbations whose wavelengths $\lambda$ are smaller  than Jeans wavelength $\lambda_J$ propagate in time as harmonic waves, since $\lambda_J/\lambda=\kappa_*>1$. For those perturbations  where $\lambda_J/\lambda<1$, there are two modes: in one the perturbation will decay in time, and in the other the perturbation will grow with time, which is referred to the Jeans instability.      

For small values of the dimensionless friction viscous coefficient, the three roots reduce to
\ben\lb{w1}
\omega_*=\pm\sqrt{\kappa_*^2-1}-i\eta_*\frac{9\kappa_*^2-5}{10(\kappa_*^2-1)},\qquad \omega_*=-i\eta_*\frac{2(3\kappa_*^2-5)}{5(\kappa_*^2-1)},
\een
where the first value above for dimensionless frequency $\omega_*$ represents a damped harmonic  wave if the perturbation wavelengths are smaller than the Jeans wavelength ($\lambda_J/\lambda>1$) otherwise a growth (instability) or decay perturbation if $\lambda_J/\lambda<1$. The second value above for $\omega_*$ refers to a perturbation that decays in time if $\lambda_J/\lambda>\sqrt{5/3}$ or a growth perturbation (instability) otherwise.

\begin{figure}[ht]
\includegraphics[width=12 cm]{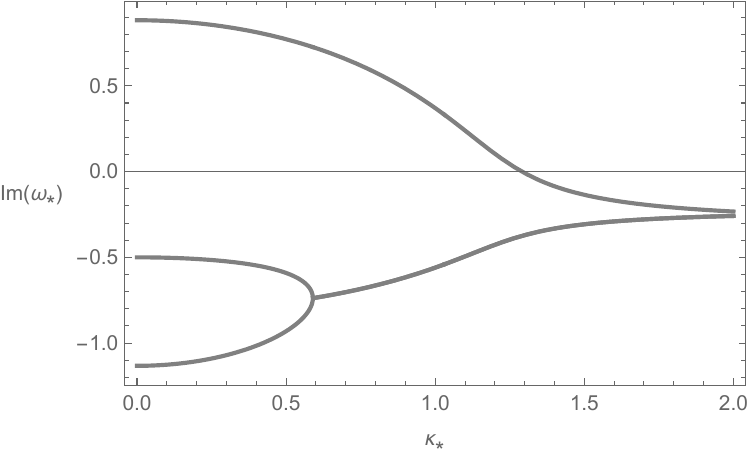}
\caption{Imaginary  part of the dimensionless frequency $\omega_*$ as function of the dimensionless wavenumber $\kappa_*$ for a dimensionless friction viscous coefficient $\eta_*=0.25$.} \label{fig2}
\end{figure}   
\unskip

The  solution of the dispersion relation (\ref{brw26a}) for the dimensionless friction viscous coefficient $\eta_*=0.25$ is represented in Figure \ref{fig1} where the real part of the dimensionless frequency $\omega_*$ is plotted as a function of the dimensionless wavenumber $\kappa_*$, while the imaginary part of $\omega_*$ is plotted in Figure \ref{fig2}.

From Figure \ref{fig1} one can infer that there are two  modes that describe harmonic wave propagation in opposite directions and one mode that does not propagate, since its real part vanishes. 

One observes from  Figure \ref{fig2} that:   (i) if ${\rm Im}[\omega_*] < 0$  the amplitude of the perturbations decays in time towards zero  and the eigenmode is  stable;  (ii) if ${\rm Im}[\omega_*] >0$  the amplitude of the perturbations growths with time and the eigenmode is unstable. Note that for large values of the dimensionless wave number $\kappa_*$ the unstable eigenmode becomes stable. 

\subsection{Post-Newtonian Fokker-Planck equation}

The  determination in the  post-Newtonian approximation  of  the  previous modes analyzed is more involved, since the Fokker-Planck equation (\ref{brw18a}) is coupled with the three Poisson equations (\ref{m2a}).

The background quantities are related to the Maxwell-J\"uttner distribution function (\ref{brw8b}) and to the gravitational potentials $U_0(\bx)$, $\Phi_0(\bx)$ and $\Pi^0_i(\bx)$. The perturbed terms are denoted by a subscript 1 and we write
\ben\lb{jpn6a}
&&f_B(\bx,\bv_B,t)=f_{MJ}^B(\bx,\bv_B,t)\left[1+  f_1(\bx,\bv_B,t)\right],
\qquad U(\bx,t)=U_0(\bx)+  U_1(\bx,t),
\\\lb{jpn6b}
&&\Phi(\bx,t)=\Phi_0(\bx)+\Phi_1(\bx,t),
\qquad
\Pi_i(\bx,t)=\Pi_i^0(\bx)+\Pi_i^1(\bx,t).
\een

The insertion of the representations (\ref{jpn6a}) and (\ref{jpn6b}) into the Fokker-Planck equation (\ref{brw18a}) follows a hierarchy of equations. The equation for the background quantities is given by
\ben\no
&&\frac{\partial U_0}{\partial x^i}\left[\frac{\partial f_{MJ}^{0B}}{\partial v_B^i}-\frac{2v_B^2f_B^{(0)}}{\sigma_B^2c^2}v^B_i\right]\left\{1+\frac1{\zeta_B}\left[\frac{15}8+\frac{v_B^2}{2\sigma_B^2}+\frac{U_0}{\sigma_B^2}+\frac{2v_B^2}{\sigma_B^2}\left(\frac{U_0}{\sigma_B^2}+\frac{3v_B^2}{16\sigma_B^2}\right)\right]\right\}
\\\lb{jpn}
&&\qquad+\frac1{\zeta_B}\left[\left(\frac{v_B^2}{\sigma_B^2}-\frac{4U_0}{\sigma_B^2}\right)\frac{\partial U_0}{\partial x^i}
-\frac{4v^B_iv^B_j}{\sigma_B^2}\frac{\partial U_0}{\partial x^j}+\frac2{\sigma_B^2}\frac{\partial \Phi_0}{\partial x^i}+\frac{v^B_j}{\sigma_B^2}\left(\frac{\partial \Pi^0_i}{\partial x^j}-\frac{\partial \Pi^0_j}{\partial x^i}\right)\right]\frac{\partial f_B^{(0)}}{\partial v_B^i}=0,
\een
where we have introduced the background Maxwell-J\"uttner distribution function $f_{MJ}^{0B}$ defined through the relationship
\ben\lb{jpn4a}
f_{MJ}^B=f_B^{(0)}
\left\{1-\frac{1}{\zeta_B}\left[\frac{15}8+\frac{3v_B^4}{8\sigma_B^4}
+2\frac{U_0 v_B^2}{\sigma_B^4}\right]\right\}-2f_B^{(0)}\frac{U_1v_B^2}{\sigma_B^4}=f_{MJ}^{0B}-2f_B^{(0)}\frac{U_1v_B^2}{\sigma_B^4}.
\een
Note that the equation for the background quantities (\ref{jpn}) is identically satisfied if $\nabla U_0=0$, $\nabla \Phi_0=0$ and $\nabla \Pi_i^0=0$. 

The equation for the perturbed quantities -- by taking into account $\nabla U_0=0$, $\nabla \Phi_0=0$ and $\nabla \Pi_i^0=0$ -- reduces to
\ben\no
&&\left\{1+\frac1{\zeta_B}\left[\frac{v_B^2}{2\sigma_B^2}+\frac{U_0}{\sigma_B^2}\right]\right\}\left[\frac{\partial f^B_1}{\partial t}+v^B_i\frac{\partial f^B_1}{\partial x^i}\right]-\frac{v_i^B}{\sigma_B^2}\frac{\partial U_1}{\partial x^i}\left\{1+\frac1{\zeta_B}\left[\frac{2v_B^2}{\sigma_B^2}+\frac{5U_0}{\sigma_B^2}\right]\right\}
\\\lb{jpn12a}
&&\qquad+\frac1{\zeta_G}\left[\frac{v_B^2}{\sigma_B^4}\left(\frac{\partial U_1}{\partial t}+v^B_i\frac{\partial U_1}{\partial x^i}\right)+\frac{4U_0}{\sigma_B^4}\frac{\partial U_1}{\partial x^i}v_i^B
-\frac{v_i^B}{\sigma_B^4}\left(\frac{\partial \Pi^1_i}{\partial t}+2\frac{\partial \Phi_1}{\partial x^i}\right)\right]=\eta\left[\sigma_B^2\frac{\partial^2 f_1^B}{\partial v_B^i\partial v_B^i}-v_i^B\frac{\partial f_1^B}{\partial v_B^i}\right].
\een

For the analysis of the equation  (\ref{jpn12a}) the perturbations are represented as plane waves of wave number vector $\bk$, frequency $\omega$ and  small amplitudes  $\overline f_1(\bv)$, $ \overline U_1$, $\overline\Phi_1$ and $\overline{\Pi^1_i}$, i.e.,
\ben\lb{jpn12b}
f_1(\bx,\bv,t)=\overline f_1(\bv)\,e^{i(\bk\cdot\bx-\omega t)},\quad U_1(\bx,t)=\overline U\,e^{i(\bk\cdot\bx-\omega t)},
\quad
\Phi_1(\bx,t)=\overline \Phi\,e^{i(\bk\cdot\bx-\omega t)},\quad \Pi_i^1(\bx,t)=\overline {\Pi_i}\,e^{i(\bk\cdot\bx-\omega t)},
\een
while the  amplitude of the distribution function $\overline f_1(\bv)$ is considered as a linear combination of the post-Newtonian summational invariants \cite{GK,GMK}, namely
\ben\lb{jpn12c}
\overline f_1(\bv)=A+ B_iv^B_i\left[1+\frac1{\zeta_B}\left(\frac{v_B^2}{2\sigma^2_B}+3\frac{U_0}{\sigma_B^2}\right)\right]+Dv_B^2\left[1+\frac1{\zeta_B}\left(\frac{3v_B^2}{4\sigma_B^2}+3\frac{U_0}{\sigma_B^2}\right)\right].
\een
Here $A,D$ and $B_i$ are  unknowns that do not depend on $\bv_B$ and $U_0$.

Insertion of the representations (\ref{jpn12b}) and (\ref{jpn12c}) into the Fokker-Planck equation for the perturbations (\ref{jpn12a}) leads to
\ben\no
&&\big(v_i^B k_i-\omega\big)\bigg\{A+ B_iv^B_i\bigg[1+\frac1{\zeta_B}\bigg(\frac{v_B^2}{2\sigma^2_B}+\frac{3U_0}{\sigma_B^2}\bigg)\bigg]+Dv_B^2\bigg[1+\frac1{\zeta_B}\bigg(\frac{3v_B^2}{4\sigma_B^2}+\frac{3U_0}{\sigma_B^2}\bigg)\bigg]\bigg\}
\\\no
&&\qquad-\frac{\overline U}{\sigma_B^2}\bigg[1+\frac1{\zeta_B}\bigg(\frac{3v_B^2}{2\sigma^2_B}+\frac{4U_0}{\sigma_B^2}\bigg)\bigg]v_i^Bk_i
+\frac1{\zeta_B}\bigg\{\frac{\overline U}{\sigma_B^2}\bigg[\bigg(\frac{v_B^2}{\sigma^2_B}+\frac{4U_0}{\sigma_B^2}\bigg)v_i^Bk_i-\frac{v_B^2}{\sigma^2_B}\omega\bigg]-\big(2k_i\overline\Phi-\omega\overline\Pi_i\big)\frac{v_i^B}{\sigma_B^2}\bigg\}
\\\lb{jpn12d}
&&\qquad +i\eta\bigg\{6D\sigma_B^2\bigg[1+\frac1{\zeta_B}\bigg(\frac{2v_B^2}{\sigma^2_B}+\frac{2U_0}{\sigma_B^2}\bigg)\bigg]\bigg[1-\frac{v_B^2}{3\sigma_B^2}\bigg(1-\frac1{\zeta_B}\frac{v_B^2}{\sigma^2_B}\bigg)\bigg]-B_iv_i^B\bigg[1+\frac1{\zeta_B}\bigg(\frac{v_B^2}{\sigma^2_B}+\frac{2U_0}{\sigma_B^2}-5\bigg)\bigg]\bigg\}.
\een

We multiply equation (\ref{jpn12d}) successively by the summational invariants and integrate the resulting equations with respect to the equilibrium distribution function, yielding
\ben\lb{s1}
&&A\omega+3D\sigma_B^2\bigg\{\bigg[1+\frac1{\zeta_B}\bigg(\frac{15}{4}+\frac{3U_0}{\sigma_B^2}\bigg)\bigg]\omega -\frac2{\zeta_B}i\eta\bigg\}-B_ik_i\sigma_B^2\bigg[1+\frac1{\zeta_B}\bigg(\frac{5}{2}+\frac{3U_0}{\sigma_B^2}\bigg)\bigg]+\frac3{\zeta_B}\frac{\overline U}{\sigma_B^2}\omega=0,
\\\no
&&A\kappa^2\bigg[1+\frac1{\zeta_B}\bigg(\frac{5}{2}+\frac{3U_0}{\sigma_B^2}\bigg)\bigg]+D\sigma_B^2\kappa^2\bigg[1+\frac1{\zeta_B}\bigg(\frac{35}{4}+\frac{6U_0}{\sigma_B^2}\bigg)\bigg]-\frac{\overline U}{\sigma_B^2}\kappa^2\bigg[1+\frac1{\zeta_B}\bigg(5+\frac{3U_0}{\sigma_B^2}\bigg)\bigg]
\\\lb{s2}
&&\qquad-B_i k_i\bigg\{\bigg[1+\frac1{\zeta_B}\bigg(5+\frac{6U_0}{\sigma_B^2}\bigg)\bigg]\omega+i\eta\bigg[1+\frac1{\zeta_B}\bigg(\frac52+\frac{5U_0}{\sigma_B^2}\bigg)\bigg]\bigg\}-\frac{2\overline\Phi}{\sigma_B^4}\frac{\kappa^2}{\zeta_B}+\frac{\overline \Pi_i k_i}{\sigma_B^4}\frac{\omega}{\zeta_B}=0,
\\\no
&&3A\omega\bigg[1+\frac1{\zeta_B}\bigg(\frac{15}4+\frac{3U_0}{\sigma_B^2}\bigg)\bigg]+15D\sigma_B^2\bigg\{\bigg[1+\frac1{\zeta_B}\bigg(\frac{7}{10}+\frac{2U_0}{5\sigma_B^2}\bigg)\bigg]\omega +\frac45i\eta\bigg[1+\frac1{\zeta_B}\bigg(10+\frac{5U_0}{\sigma_B^2}\bigg)\bigg]\bigg\}
\\\lb{s3}
&&\qquad-5B_ik_i\sigma_B^2\bigg[1+\frac1{\zeta_B}\bigg(\frac{35}{4}+\frac{6U_0}{\sigma_B^2}\bigg)\bigg]+\frac{15}{\zeta_B}\frac{\overline U}{\sigma_B^2}\omega=0.
\een

The system of equations (\ref{s1}) -- (\ref{s3}) is a function of the amplitudes $A, D, B_ik_i, \overline U, \overline\Phi, \overline \Pi_i k_i $ and to have an algebraic system for these amplitudes we have to consider the Poisson equations (\ref{m2a}), which follow from the knowledge of the energy-momentum tensor components.

For the determination of the energy-momentum tensor components, we write the four-velocity components  and the invariant integration element as
\ben\lb{s4}
u^0_B=c\left[1+\frac1{\zeta_B}\left(\frac{v_B^2}{2\sigma_B^2}+\frac{U_0+ U_1}{\sigma_B^2}\right)\right],\quad u_B^i=\frac{u_B^0v_B^i}c,\quad
\frac{\sqrt{-g}\, d^3 u_B}{u^B_0}=
\left\{1+\frac1{\zeta_B}\left[\frac{2v^2+6U_0+6 U_1}{\sigma_B^2}\right]\right\}\frac{d^3v_B}c.
\een
By inserting the representation of the one-particle distribution function of the Brownian particles (\ref{jpn6a}) together with (\ref{jpn4a}), (\ref{jpn12b}) and (\ref{s4}) into the definition of the energy-momentum tensor and subsequent integration of the resulting equation, we get 
\ben
&&{\buildrel\!\!\!\! _{0} \over{T_B^{00}}}=c^2\rho_B^{(0)}\left[1+e^{i(\bk\cdot\bx-\omega t)}\left(A+D\sigma^2_B\right)\right],
\\
&&{\buildrel\!\!\!\! _{2} \over{T_B^{00}}}=\rho_B^{(0)}\sigma_B^2\left\{\frac32+\frac{2U_0}{\sigma_B^2}+e^{i(\bk\cdot\bx-\omega t)}\left[A\left(\frac32+\frac{2U_0}{\sigma_B^2}\right)+3D\sigma^2_B\left(\frac{15}4+\frac{U_0}{\sigma_B^2}\right)+\frac{2\overline U}{\sigma_B^2}\right]\right\},
\\
&&{\buildrel\!\!\!\! _{2}\over{T_B^{ii}}}=3\rho_B^{(0)}\sigma_B^2\left[1+e^{i(\bk\cdot\bx-\omega t)}\left(A+5D\sigma^2_B\right)\right],\qquad
{\buildrel\!\!\!\! _{1}\over{T_B^{0i}}}=c\rho_B^{(0)}\,e^{i(\bk\cdot\bx-\omega t)}\sigma_B^2B_i.
\een

Here we invoke again the "Jeans swindle" and consider that the Poisson equations (\ref{m2a}) are valid only for the perturbed quantities and that they refer only to the Brownian particles, due to the approximation $m_G/m_B\ll1$. Hence, we have
\ben\lb{s5}
&&\kappa_*^2\frac{\overline U}{\sigma_B^2}=A+3\sigma_B^2D,\qquad \kappa_*^2\frac{\overline \Pi_ik_i}{\sigma_B^4}=4B_ik_i-\frac{\kappa^2}{\sigma_B^2}\omega\frac{\overline U}{\sigma_B^2},
\\\lb{s6}
&&2\kappa_*^2\frac{\overline \Phi}{\sigma_B^4}=A\left(\frac92+\frac{2U_0}{\sigma_B^2}\right)+3D\sigma^2_B\left(\frac{35}4+\frac{U_0}{\sigma_B^2}\right)+\frac{2\overline U}{\sigma_B^2}.
\een

Now the system of equations (\ref{s1}) -- (\ref{s3}) together with (\ref{s5}) and (\ref{s6}) becomes an algebraic system of equations for the determination of the amplitudes $A, D, B_ik_i, \overline U, \overline\Phi, \overline \Pi_i k_i $. This coupled system of equations has a non-trivial solution of this coupled system  if the determinant of their  coefficients vanishes, which implies the dispersion relation
\ben\no
&&\omega_*\left(\omega_*^2-\kappa_*^2+1-2\eta_*^2\right)+i\eta_*\left(2+3\omega_*^2-\frac65\kappa_*^2\right)+\frac1{\zeta_B}\bigg\{\omega_*\bigg[\frac{55}{4}+10\bigg(\frac{U_0}{\sigma_B^2}+\frac1{3\kappa_*^2}\bigg)-\kappa_*^2\bigg(22+\frac{12U_0}{\sigma_B^2}\bigg)-\bigg(28+\frac{20U_0}{\sigma_B^2}
\\
&&+\frac{10}{\kappa_*^2}\bigg)\eta_*^2\bigg]+\omega_*^3\bigg(20+\frac{12U_0}{\sigma_B^2}\bigg)+i\eta_*\bigg[49+\frac{26U_0}{\sigma_B^2}+\frac{20}{3\kappa_*^2}+\omega_*^2\bigg(\frac{101}2+\frac{33U_0}{\sigma_B^2}+\frac{5}{\kappa_*^2}\bigg)-\kappa_*^2\bigg(21+\frac{66U_0}{5\sigma_B^2}\bigg)\bigg]\bigg\}=0.
\een
In the above equation, we have considered the solution in the first post-Newtonian approximation, i.e., up to the relativistic $1/\zeta_B$-term. 

For small values of the the dimensionless friction viscous coefficient $\eta_*$, we get three roots for the dimensionless frequency $\omega_*$ where the relativistic corrections $1/\zeta_B$ to the solutions (\ref{w1}) show up, namely
\ben\no
&&\omega_*=\pm\sqrt{\kappa_*^2-1}-i\eta_*\frac{9\kappa_*^2-5}{10(\kappa_*^2-1)}+\frac1{\zeta_B}\bigg\{\frac{\pm3\kappa_*^2(4\kappa_*^2+4U_0/\sigma_B^2-15)\mp20}{12\kappa_*^2\sqrt{\kappa_*^2-1}}
\\
&&\qquad+i\eta_*\frac{3\kappa_*^6(41+18U_0/\sigma_B^2)-6\kappa_*^4(85+46U_0/\sigma_B^2)+5\kappa_*^2(65+54U_0/\sigma_B^2)-150}{60\kappa_*^2(\kappa_*^2-1)}\bigg\},
\\ 
&&\omega_*=-i\eta_*\left\{\frac{2(3\kappa_*^2-5)}{5(\kappa_*^2-1)}+\frac1{\zeta_B}\frac{130-150\kappa_*^2+81\kappa_*^4+6U_0/\sigma_B^2(3\kappa^4_*+8\kappa_*^2-15)}{15(\kappa_*^2-1)^2}\right\}.
\een
From the solutions given above we note that in the  relativistic correction $1/\zeta_B$, there exists a dependence of the dimensionless frequency $\omega_*$  on the background Newtonian gravitational potential $U_0$. This dependence was also reported in  some works on the Jeans stability \cite{GMK,NKR,NH,GK,GK1,GK2}.

\section{Conclusions}

In this work, we have determined  the Fokker-Planck equation for the Brownian motion within the framework of the post-Newtonian Boltzmann equation. The system analyzed was a mixture of  a perfect gas of light rest mass particles and  Brownian particles with heavy rest mass, where the  particle number density  of the Brownian particles is much smaller than the one of the perfect gas. At equilibrium, both species were characterized by post-Newtonian Maxwell-J\"uttner distribution functions. The collision term of the pos-Newtonian Boltzmann equation of the Brownian particles was described only by the collisions between the two species, since  the collisions between the Brownian particles were neglected due to the assumption of small particle number density. In the non-relativistic regimes, where $m_Bc^2/kT\gg1$ and $m_Gc^2/kT\gg1$  the post-Newtonian Fokker-Planck equation for the Brownian motion and the friction viscous coefficient reduce to the expressions given in the work by Wang-Chang and Uhlenbeck \cite{WCU}. The relativistic correction  for the  friction viscous coefficient depends on the Newtonian gravitational potential $U$ in the first post-Newtonian approximation and in the second post-Newtonian approximation on the gravitational potentials $\Phi, \Psi_{kk}$. Without the gravitational potentials the expression given in \cite{Guilh} for the relativistic friction viscous coefficient was recovered. The linear stability of the Newtonian and post-Newtonian Fokker-Planck equations for the Brownian motion was analyzed by considering the perturbations from a background state as plane harmonic waves of small amplitudes. From a dispersion relation three modes show up: for perturbation wavelengths smaller than Jeans wavelength  two propagating modes that describe harmonic wave propagation in opposite directions and one mode that does not propagate. For perturbation wavelengths that are larger than the Jeans wavelength, the  time evolution corresponds  to a decay   or a growth of the perturbation; the one which grows refers to an instability.

\acknowledgments{ Conselho Nacional de Desenvolvimento Cient\'{i}fico e Tecnol\'{o}gico (CNPq), Brazil (grant No.  304054/2019-4).}

\end{document}